\documentclass[sigconf]{acmart}
\AtBeginDocument{%
  }

\usepackage{enumitem}
\usepackage{bm}
\usepackage{booktabs}
\usepackage{xcolor}
\usepackage{colortbl}

\usepackage{fancyvrb}

\copyrightyear{2026}
\acmYear{2026}
\acmDOI{XXXXXXX.XXXXXXX}
\acmConference[SAC'26]{The 41st ACM/SIGAPP Symposium on Applied Computing}{March 23--27, 2026}{Thessaloniki, Greece}
\acmISBN{979-X-XXXX-XXXX-X/26/03}

\begin{document}

\pagenumbering{gobble}
\setcounter{page}{0}

\section*{ACM Copyright Statement}

Copyright \textcopyright 2026 held by the owner/author(s). Publication rights licensed to ACM. Permission to make digital or hard copies of all or part of this work for personal or classroom use is granted without fee provided that copies are not made or distributed for profit or commercial advantage and that copies bear this notice and the full citation on the first page. Copyrights for components of this work owned by others than the author(s) must be honored. Abstracting with credit is permitted. To copy otherwise, or republish, to post on servers or to redistribute to lists, requires prior specific permission and/or a fee. Request permissions from permissions@acm.org.

\vspace{3em}
\noindent\textbf{Published in:} Proceedings of the 41st ACM/SIGAPP Symposium On Applied Computing, 2026 (\textit{SAC'26, Thessaloniki, Greece})

\vspace{3em}
\noindent\textbf{Cite as:}
\vspace{0.5em}

\noindent
Pedro R. Pires, Gregorio F. Azevedo, Rafael T. Sereicikas, Pietro L. Campos, and Tiago A. Almeida. 2026. The Bandit's Blind Spot: The Critical Role of User State Representation in Recommender Systems. In \textit{Proceedings of the 41st ACM/SIGAPP Symposium On Applied Computing} (Thessaloniki, Greece) \textit{(SAC'26)}. Association for Computing Machinery, New York, NY, USA, 1190-1197, doi:10.1145/3748522.3779905

\vspace{3em}
\noindent\textbf{BibTeX:}
\vspace{0.5em}

\begin{Verbatim}[frame=single]
@inproceedings{10.1145/3748522.3779905,
    title       =   {The Bandit's Blind Spot: The Critical Role of User State Representation in Recommender 
                     Systems},
    author      =   {Pedro R. Pires and Gregorio F. Azevedo and Rafael T. Sereicikas and Pietro L. Campos and
                     Tiago A. Almeida},
    booktitle   =   {Proceedings of the 41st ACM/SIGAPP Symposium On Applied Computing},
    series      =   {SAC'26},
    location    =   {Thessaloniki, Greece},
    publisher   =   {Association for Computing Machinery},
    address     =   {New York, NY, USA},
    pages       =   {1190--1197},
    year        =   {2026},
    doi         =   {10.1145/3748522.3779905},
    keywords    =   {recommender systems, contextual multi-armed bandits, state representation, embeddings,
                     matrix factorization, incremental learning},
}
\end{Verbatim}

\newpage
\pagenumbering{arabic}

\title{The Bandit's Blind Spot: The Critical Role of User State Representation in Recommender Systems}

\renewcommand{\shorttitle}{The Critical Role of User State Representation in Recommender Systems}

\author{Pedro R. Pires}
    \authornote{Corresponding author}
    \orcid{0000-0001-7990-9097}
    \affiliation{%
        \institution{Federal University of São Carlos}
        \streetaddress{Rod. João Leme dos Santos km 110}
        \city{Sorocaba} 
        \state{São Paulo} 
        \country{Brazil}
        \postcode{18052-780}  
    }
    \email{pedro.pires@dcomp.sor.ufscar.br}

\author{Gregorio F. Azevedo}
    \orcid{0000-0002-1096-7456}
    \affiliation{%
        \institution{Federal University of São Carlos}
        \streetaddress{Rod. João Leme dos Santos km 110}
        \city{Sorocaba} 
        \state{São Paulo} 
        \country{Brazil}
        \postcode{18052-780}  
    }
    \email{gregorio.fornetti@estudante.ufscar.br}

\author{Rafael T. Sereicikas}
    \orcid{0009-0009-9198-5469}
    \affiliation{%
        \institution{Federal University of São Carlos}
        \streetaddress{Rod. João Leme dos Santos km 110}
        \city{Sorocaba} 
        \state{São Paulo} 
        \country{Brazil}
        \postcode{18052-780}  
    }
    \email{rafaeltofoli@estudante.ufscar.br}

\author{Pietro L. Campos}
    \orcid{0009-0004-3378-5921}
    \affiliation{%
        \institution{Federal University of São Carlos}
        \streetaddress{Rod. João Leme dos Santos km 110}
        \city{Sorocaba} 
        \state{São Paulo} 
        \country{Brazil}
        \postcode{18052-780}  
    }
    \email{pietro.campos@estudante.ufscar.br}

\author{Tiago A. Almeida}
    \orcid{0000-0001-6943-8033}
    \affiliation{%
        \institution{Federal University of São Carlos}
        \streetaddress{Rod. João Leme dos Santos km 110}
        \city{Sorocaba} 
        \state{São Paulo} 
        \country{Brazil}
        \postcode{18052-780}  
    }
    \email{talmeida@ufscar.br}

\renewcommand{\shortauthors}{P. R. Pires et al.}

\begin{abstract}
    With the increasing availability of online information, recommender systems have become an important tool for many web-based systems. Due to the continuous aspect of recommendation environments, these systems increasingly rely on contextual multi-armed bandits (CMAB) to deliver personalized and real-time suggestions. A critical yet underexplored component in these systems is the representation of user state, which typically encapsulates the user's interaction history and is deeply correlated with the model's decisions and learning. In this paper, we investigate the impact of different embedding-based state representations derived from matrix factorization models on the performance of traditional CMAB algorithms. Our large-scale experiments reveal that variations in state representation can lead to improvements greater than those achieved by changing the bandit algorithm itself. Furthermore, no single embedding or aggregation strategy consistently dominates across datasets, underscoring the need for domain-specific evaluation. These results expose a substantial gap in the literature and emphasize that advancing bandit-based recommender systems requires a holistic approach that prioritizes embedding quality and state construction alongside algorithmic innovation. The source code for our experiments is publicly available on \url{https://github.com/UFSCar-LaSID/bandits_blind_spot}.
\end{abstract}




\maketitle

\section{Introduction}

The explosive growth of online information has made recommender systems a key component of modern web platforms, helping users navigate catalogs of music, videos, products, and other media. By leveraging past interactions, these systems learn user preferences, filter relevant items, and deliver personalized suggestions~\cite{BobadillaSurvey2013}.

Early recommenders operated in a static setting: algorithms were trained on a fixed batch of historical data and retained this knowledge unchanged during inference, being retrained from scratch when necessary~\cite{GamaEvaluating2012}. Such an approach overlooks a defining characteristic of real-world recommendation environments: the continuous stream of user interactions and feedback. In practical web-based systems, a user receives a recommendation and immediately provides feedback. Incremental recommenders can then incorporate this novel information to update their models~\cite{VinagreOverview2015}.

One of the most established families of incremental algorithms is the multi-armed bandit framework~\cite{SilvaMulti2022}. Bandits operate as sequential decision-making models, selecting actions and receiving rewards that are then used to refine future decisions. Contextual multi-armed bandits (CMABs) have become a key approach for generating recommendations in real-time systems. They encode users and items as feature vectors to represent the current state, select an item (\textit{arm}) to recommend, and collect feedback (\textit{reward}) from user interactions---such as clicks, views, or purchases---to update the underlying model incrementally. A foundational CMAB variant is the linear bandit, which scores items using a linear regression model and applies an exploration strategy to balance the exploration–exploitation trade-off~\cite{ChenValue2021}.

In the following years, the literature on CMAB-based recommenders expanded considerably. Numerous strategies have been explored to enhance performance, including pre-filtering items to reduce the action space~\cite{JagermanWhen2019,LiuTransferable2018}, clustering users or items to generalize learning~\cite{CelisControlling2019,WangFactorization2017}, incorporating deep learning or graph-based models~\cite{KakadiyaRelational2020,QiGraph2023}, and employing structured bandits to recommend slates of items~\cite{HwangCombinatorial2023,LiCascading2020}. Broadly, most research has concentrated on four key aspects of MABs: \textit{(i)} the underlying model that consumes the user state; \textit{(ii)} the exploration strategy that balances exploration and exploitation; \textit{(iii)} the inference procedure, i.e., how the final recommendation is produced; and \textit{(iv)} the reward modeling, which defines both the model learning and the evaluation protocol.

However, current literature in the field often neglects a critical aspect of CMABs: the \textbf{state representation}~\cite{GanKnowledge2022}. This contextual representation typically encodes the user's interaction history, which may be represented through a simple one-hot encoded vector, a summary of consumed item metadata, or---more commonly---user and item embeddings~\cite{HejaziniaAccelerated2019}. These embeddings can be learned during training or obtained from external sources such as matrix factorization models. Yet, most CMAB studies treat state representation as granted~\cite{LiCollaborative2016,ZhuScalable2023}, adopting one of these conventional strategies without investigating how this seemingly minor design choice can substantially influence the model's final performance.

In this paper, we demonstrate how the state representation can deeply impact the accuracy of the model. By training static user and item embeddings and exploring multiple formats for incorporating them into the model input, we reveal that final performance can vary significantly depending on the chosen representation. Moreover, our results reveal no clear winner among representation strategies, highlighting the need to evaluate multiple context representations per application domain.

In summary, our main contributions are:
\begin{itemize}[leftmargin=*, itemsep=2pt]
    \item[\textit{(i)}] To the best of our knowledge, we present the first large-scale empirical comparison of state representations in contextual multi-armed bandits, evaluating six widely used recommendation data\-sets, two matrix factorization models, three distinct representation strategies and three CMAB models. Unlike prior work, we systematically vary both the embedding method and the representation formulation to isolate their impact on bandit performance;
    \item[\textit{(ii)}] We show that the choice of bandit algorithm often results in minor performance differences compared to the substantial impact of the selected context representation strategy;
    \item[\textit{(iii)}] We identify a prominent gap in the literature: while much research emphasizes developing new algorithms, reward modeling techniques, and evaluation methods, the design of the environment's state representation remains largely overlooked.
\end{itemize}

\section{Related Work}

Incremental learning has become essential for recommender systems operating in dynamic environments~\cite{GamaEvaluating2012}. A fundamental building block of incremental recommender systems is the contextual multi-armed bandit framework (CMABs), which leverages user feedback to refine its recommendations~\cite{SilvaMulti2022}. Exemplified by methods such as LinGreedy~\cite{LangfordEpoch2007} and LinUCB~\cite{LiContextual2010}, CMABs extend traditional bandits by incorporating contextual state representations, allowing incremental algorithms to leverage environmental information when making recommendations.

Although the defining feature of CMABs is their use of contextual information, most research has predominantly focused on reward modeling and algorithmic improvements. In contrast, relatively little attention has been given to the design of effective user state representations~\cite{GanKnowledge2022}. Broadly, there are two main strategies for generating context in MAB-based environments: \textit{(i)} leveraging user and item content-based features, and \textit{(ii)} employing embeddings that represent these entities.

When using content-based representations, CMAB models typically characterize contexts through user demographic information and item metadata. This information is often readily available within the employed datasets and is particularly useful for addressing cold-start problems and enhancing explainability~\cite{McInerneyExplore2018}. Feature vectors are used directly as input to bandit models~\cite{GuoDeep2020} or undergo some form of transformation to better adapt to the task before being consumed~\cite{ZhuScalable2023}. However, content-based user and item features are not always available and are often subject to noise and inconsistencies, which can limit the applicability of such systems~\cite{PazzaniContent2007}.

Due to the limitations of explicit features, a common strategy is to represent users and items using latent embeddings. These low-dimensional vectors capture user preferences and item characteristics within a shared latent space~\cite{KorenMatrix2009}. Examples include the warm-up method of~\citep{MaryBandits2014}, which employs Alternating Least Squares to generate embeddings used to initialize bandits; Latent Contextual Bandits (LCB)~\cite{ZhouLatent2016} which leverages user latent classes---closely related to user embeddings---to provide recommendations for cold-start users; BanditMF~\cite{XuBanditMF2021}, a MAB-based model that utilizes MF embeddings to cluster users and guide a bandit module; and the healthcare-focused recommender of~\citet{ZhouSpoiled2023}, which learns item and user embeddings through a deep learning approach and concatenates them as input to the bandit algorithm.

Many existing recommenders that use item and user embeddings as context rely on offline pre-trained embeddings that remain static during online user interactions. While this approach simplifies modeling, it may fail to capture the evolving dynamics inherent in real-time recommender systems. To address this limitation, several studies focus on learning evolving embeddings. Examples include COFIBA~\cite{LiCollaborative2016}, which combines LinUCB with user clustering by dynamically grouping users based on item feature vectors; Alternating Linear Bandits~\cite{DadkhahiAlternating2018}, which continuously retrains MF-based embeddings used as context; and the work of~\citet{KhosraviPreferences2024}, which explores evolving user states and their representation.

Even when accounting for the dynamic nature of the problem, the vast majority of studies tend to pay limited attention to how item and user feature vectors are obtained and how state representations are constructed, placing far greater emphasis on the bandit algorithms themselves. In many works, the context is treated as given, without any mention of how to build it~\cite{LiCollaborative2016,ZhuScalable2023, FeijerCalibrated2025, GuoUncertainty2024}.

However, studies that prioritize state representation tend to achieve more impactful results. For instance, \citet{HeDynamic2023} dynamically learned variable-sized embeddings within a bandit-based framework to feed a neural model, outperforming baseline methods. More recently, \citet{FengContextual2024} and \citet{CuiContext2024} demonstrated that refining representations for sequential recommendation, via bandit-based denoising and diffusion-enhanced contrastive learning, respectively, leads to significant performance improvements. Similarly, the findings of~\citet{GanKnowledge2022}, which employed knowledge graph-based embeddings to better represent items and users, demonstrate that decisions regarding state representation can have a greater impact than the choice of algorithm itself.

Motivated by this gap, our study conducts a large-scale comparison of different embedding-based algorithms and aggregation strategies for representing context in CMAB models. While many studies compare CMABs, highlighting their respective strengths and characteristics~\cite{ZhuComparative2025}, ours focus primarily on state representation. A related study by~\citet{HejaziniaAccelerated2019} examined deployment aspects and compared different types of embeddings, but did not focus on the bandit framework.

\section{Experimental Setup}

In this section, we detail the experimental setup. First, we present the datasets used in the experiments, then we describe the evaluation framework and the strategies for representing user state through static embeddings.

\subsection{Datasets}

Our experimental evaluation relies on a set of benchmark datasets frequently adopted in recommender systems and contextual bandit studies, summarized in Table~\ref{tab:datasets}. They are all publicly available and correspond to different domains. All datasets underwent a preprocessing stage to ensure consistency, which involved removing duplicate interactions and filtering out contradictory records (e.g., identical user–item pairs with different ratings).

\begin{table}[htpb]
\centering
\caption{Datasets used in the experiments.}
\label{tab:datasets}
\begin{tabular}{lccc}
\toprule
\textbf{Dataset} & \textit{\#users} & \textit{\#items} & \textit{\#interactions}\\
\midrule
\textbf{Amazon Beauty}\footnotemark[1]          &    631,986 &  112,565 &  701,528\\
\textbf{Amazon Books}\footnotemark[1]        &  1,008,954 &  206,710 &  2,437,999\\
\textbf{BestBuy}\footnotemark[2]        &  1,268,702 &  69,858 &  1,865,269\\
\textbf{Delicious}\footnotemark[3]        &  1,867 &  69,198 &  437,593\\
\textbf{MovieLens-100K}\footnotemark[4]        &    943 &  1,682 &    100,000\\
\textbf{MovieLens-25M}\footnotemark[4]      &     162,541 &  59,047 &   25,000,095\\
\textbf{RetailRocket}\footnotemark[5]      &     1,407,580 &   235,061 &    2,755,641\\
\bottomrule
\end{tabular}
\end{table}

\footnotetext[1]{Amazon Reviews'23. Available at: \url{https://amazon-reviews-2023.github.io} (visited on \today)}
\footnotetext[2]{Data Mining Hackathon on BIG DATA (7GB). Available at: \url{www.kaggle.com/c/acm-sf-chapter-hackathon-big} (visited on \today)}
\footnotetext[3]{HetRec2011 | GroupLens. Available at: \url{https://grouplens.org/datasets/hetrec-2011/} (visited on \today)}
\footnotetext[4]{MovieLens | GroupLens. Available at: \url{https://grouplens.org/datasets/movielens/} (visited on \today)}
\footnotetext[5]{Retailrocket recommender system dataset. Available at \url{https://www.kaggle.com/datasets/retailrocket/ecommerce-dataset} (visited on \today)}

\subsection{Experimental protocol}

\begin{figure*}
    \centering
    \includegraphics[width=\linewidth]{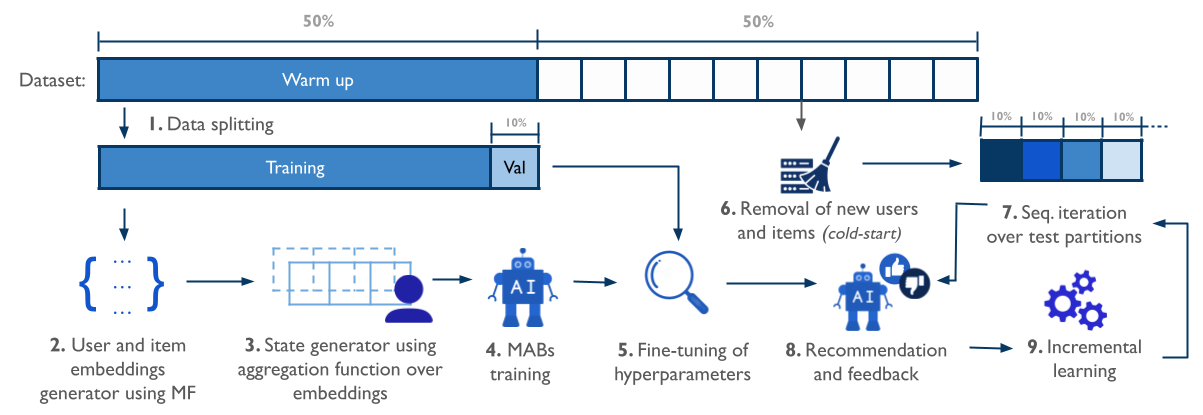}
    \caption{Experimental protocol used to benchmark the models.}
    \label{fig:experimental-protocol}
    \Description[Diagram of experimental protocol]{Blue-colored sequential diagram showing the adopted experimental protocol. The pipeline is divided into 9 steps, each with a corresponding icon, illustrating what was described in the text.}
\end{figure*}

The evaluation followed a continuous protocol, outlined in Figure~\ref{fig:experimental-protocol}. Interaction logs were ordered by timestamp to reproduce a temporal recommendation flow. From each dataset, the first 50\% of the interactions were designated as historical data for model initialization. Within this portion, a 10\% slice was set aside for hyperparameter tuning, while the remainder was used to learn the embedding representation for users and items, and warm-up the CMABs algorithms. The second half of the data was dedicated exclusively to online evaluation.

User contexts were derived by multiple types of aggregation of pre-computed user and item embeddings. These contexts served as input to the contextual bandit models, whose parameters were optimized with respect to the NDCG@20 on the validation split.

Since the embedding space was generated in a single, non-in\-cremental pass, unseen items in the test set lacked corresponding vector representations. To ensure a consistent evaluation, we excluded interactions involving those items, thereby avoiding the cold-start issue. We also chose not to employ dynamically learned embeddings in order to prevent effects related to temporal drift and model coupling. This design allowed us to isolate the influence of state representation and enable a more controlled comparison.

For the online stage, the test data was then processed in 10 equally-sized chronological windows, each containing 10\% of the interactions. In each window, the models generated ranked recommendations, observed the actual user choice, and updated their linear predictors to incorporate the new feedback.

The experimental pipeline was implemented in \texttt{Python3}, using the \texttt{implicit} library for generating the embeddings\footnotemark[6] and \texttt{Mab2Rec} library~\cite{KadiogluMab2Rec2024} to train the models.

\footnotetext[6]{Implicit -- Fast Python Collaborative Filtering for Implicit Datasets. Available at: \url{https://github.com/benfred/implicit} (visited on \today)}

\subsection{Embedding-based states}\label{sec:emb-state}

The embedding space was generated using two matrix factorization methods: \textbf{implicit Alternating Least Squares} (ALS)~\cite{HuCollaborative2008} and \textbf{Bayesian Personalized Ranking} (BPR)~\cite{RendleBPR2009}. Both approaches leverage historical user–item interactions, but differ in their optimization objectives---ALS incorporates confidence-weighted feedback, whereas BPR optimizes a ranking-based loss. These models were selected because they are well-established baselines in the literature and can be trained efficiently at scale.

Matrix factorization models represent user preferences and item characteristics in a shared latent space~\cite{KorenMatrix2009}. Given a set of observed interactions between users $u \in \mathcal{U}$ and items $i \in \mathcal{I}$, the goal is to approximate the interaction signal $r_{u,i}$ (e.g., an implicit feedback value) as the dot product between a user feature vector $\mathbf{p}_u \in \mathbb{R}^d$ and an item feature vector $\mathbf{q}_i \in \mathbb{R}^d$, such that

\begin{equation}
    r_{ui} \approx \mathbf{p}_u^\top \mathbf{q}_i
\end{equation}

Here, $d$ denotes the dimensionality of the latent space, and the matrices $\mathbf{P} \in \mathbb{R}^{|\mathcal{U}| \times d}$ and $\mathbf{Q} \in \mathbb{R}^{|\mathcal{I}| \times d}$ collect the user and item feature vectors, respectively. 

The learning process consists of adjusting $\mathbf{P}$ and $\mathbf{Q}$ to minimize a loss function defined over the observed interactions. Once trained, these feature vectors provide low-dimensional representations of users and items that can be leveraged as input for numerous downstream applications, including feeding recommendation models.

For both matrix factorization models, hyperparameters were tuned using the validation split of the warm-up data. We searched over the number of latent factors $d$, learning rate $\alpha$ (when applicable), regularization strength $\lambda$, and number of training epochs $C$. For each dataset, the configuration yielding the best NDCG@20 was selected, and the resulting embedding space was used in all subsequent bandit experiments. The full hyperparameter search ranges are summarized in Table~\ref{tab:embedding-hyperparams}.

\begin{table}[htpb]
\centering
\caption{Hyperparameter search ranges used during the embedding fine-tuning stage.}
\label{tab:embedding-hyperparams}
\begin{tabular}{lcc}
\toprule
\textbf{Parameter} & \textbf{ALS} & \textbf{BPR} \\
\midrule
$d$ & \texttt{[$4$, $8$, $16$, $32$]} & \texttt{[$4$, $8$, $16$, $32$]} \\
$\alpha$ & - & \texttt{[$10^{-3}$, $10^{-2}$, $10^{-1}$]} \\
$\lambda$ & \texttt{[$10^{-3}$, $10^{-2}$, $10^{-1}$]} & \texttt{[$10^{-3}$, $10^{-2}$, $10^{-1}$]}\\
$C$ & \texttt{[$5$, $15$, $30$]} & \texttt{[$50$, $100$, $150$]} \\
\bottomrule
\end{tabular}
\end{table}

We then explored three distinct aggregation strategies to construct the user state representation. These strategies are detailed in the following subsections.

\subsubsection{\textbf{User embeddings}}

The first and most straightforward approach is to represent the state solely by the user embedding $\mathbf{p}_u$ when generating recommendations for user $u$. This vector corresponds directly to the target user, offering the most specific and personalized representation available in the embedding space.

The main limitation of this strategy stems from the static nature of the embeddings: once computed, $\mathbf{p}_u$ remains unchanged throughout the online evaluation. Consequently, even when a user interacts with new items, their state representation stays fixed, and any variation in the recommendations arises solely from updates in the underlying linear model or the bandit exploration strategy. This results in stable but little-diversified recommendations, as the input context remains constant across rounds.

Such a representation can be effective in domains where user preferences are relatively stable. However, in dynamic environments prone to concept drift---where preferences shift rapidly---this static representation fails to capture behavioral changes, potentially degrading recommendation quality over time. 

One way to address this limitation in dynamic environments, while still relying on static embeddings, is to derive the user state from the embeddings of recently consumed items.

\subsubsection{\textbf{Item-averaged embeddings}}

This strategy constructs the user state by averaging the embeddings of all previously consumed items by user $u$. Formally, the state is computed as

\begin{equation}
    \text{state}_u = \frac{1}{|\mathcal{I}_u|} \sum_{i \in \mathcal{I}_u} \mathbf{q}_i,
\end{equation}

\noindent where $\mathcal{I}_u$ denotes the set of items consumed by $u$.  

Compared to relying solely on static user embeddings, this strategy dynamically adapts to changes in user behavior by updating the state representation as new items are consumed. Additionally, the embeddings' fixed dimensionality $d$ is maintained, facilitating efficient model training and inference.    

However, this approach does not encode the sequential order of interactions. While limiting the aggregation to the most recent items can partially preserve recency effects, the temporal sequence itself is lost during averaging. To address this limitation, alternative aggregation methods, such as the concatenation of embeddings, can be employed to retain some notion of item ordering.

\subsubsection{\textbf{Concatenated item embeddings}}

To better preserve the sequential information lost in averaging, we consider an aggregation strategy based on concatenating the embeddings of the $h$ most recent items consumed by the user, following a \textit{first-in-first-out} order. This approach explicitly encodes the order of interactions by queuing item embeddings into a longer vector, allowing the model to differentiate between recent and older interactions. Formally, the user state at time $t$ is constructed as:

\begin{equation}
    \text{state}_u^{(t)} = \bigl[\mathbf{q}_{i_{t}}, \mathbf{q}_{i_{t-1}}, \ldots, \mathbf{q}_{i_{t-h+1}}\bigr] \in \mathbb{R}^{d \times h},
\end{equation}

\noindent where \(\mathbf{q}_{i_{t}}\) is the embedding of the most recently consumed item, and the concatenation proceeds backward through the last \(h\) items in the user’s interaction history. If fewer than \(h\) items are available, zero-padding is used to maintain a consistent input size.
         
By retaining temporal order, concatenation allows contextual bandit models to capture richer user dynamics and potentially improve recommendations in scenarios with evolving preferences. However, this comes at the cost of increased dimensionality, increasing the computational and memory complexity of the algorithms linearly and quadratically with $h$, respectively. Unlike the other state representation methods, this approach requires more computational resources as the number of $h$ increases. In that way, even though this state representation can lead to better recommendations, this method has the biggest computational complexity among the tested state representation strategies, which could make its use unfeasible in some systems.

\section{Results and Discussions}

We conducted a set of experiments following the previously mentioned experimental protocol to address the following central research questions (RQs) on the evaluation of state representation for bandit recommendation:

\begin{itemize}[leftmargin=*, wide, labelwidth=!, labelindent=0pt, itemsep=3pt]
    \item \textbf{RQ1:} How do different embedding models compare in their ability to capture user–item relevance under a multi-armed bandit recommendation setting?
    \item \textbf{RQ2:} How does the choice of the embedding usage strategy (e.g., static user embeddings, aggregated item embeddings, concatenated recent items) influence evaluation outcomes?
    \item \textbf{RQ3:} How does the quality of the state representation affect performance relative to the choice of bandit algorithm used for action selection?
\end{itemize}

\begin{figure*}[!hptb]
\centering
\includegraphics[width=0.75\textwidth]{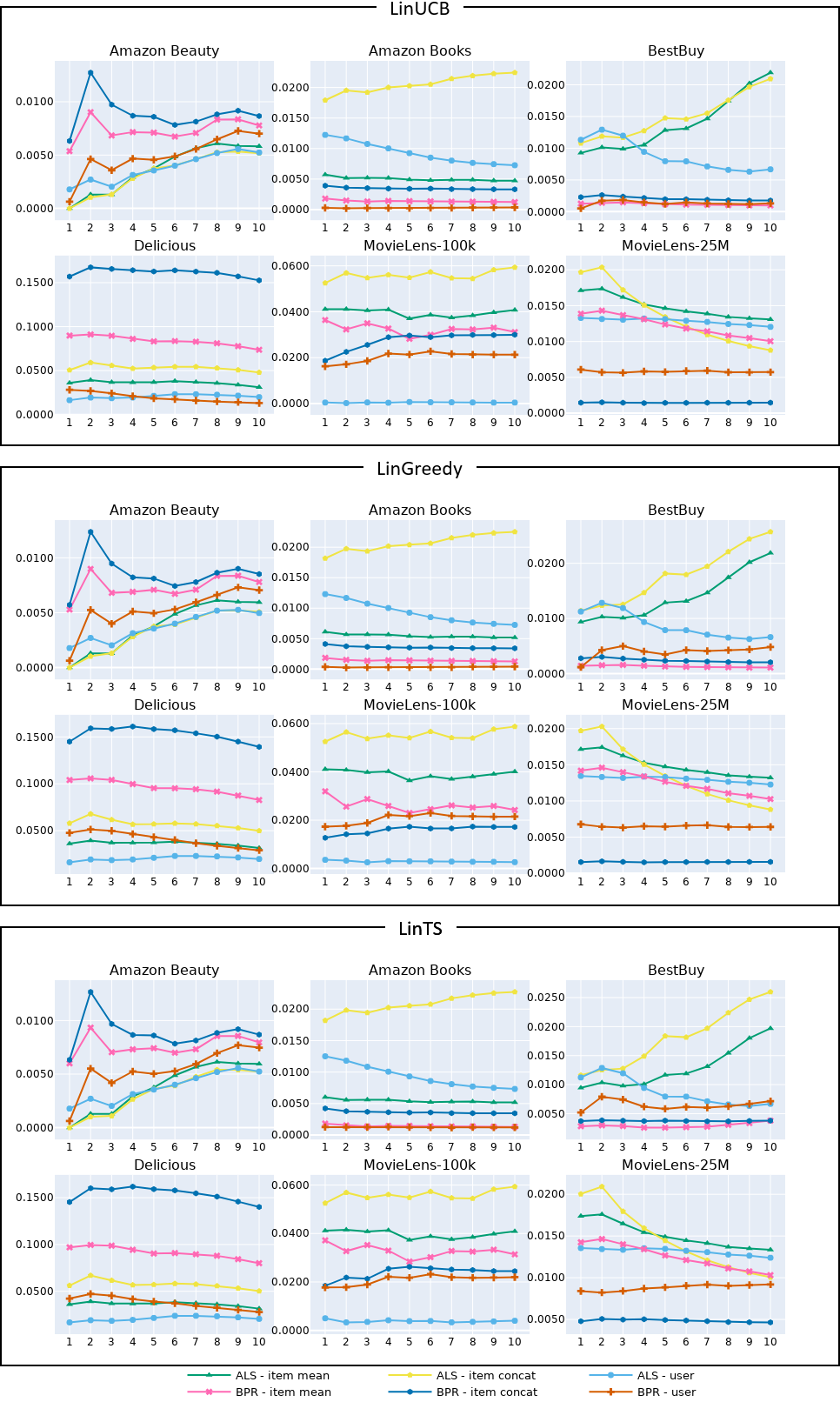}
\caption{Cumulative NDCG@20 for every partition on the test set.}
\label{fig:ndcg_results}
\Description[Grid of 18 line plots for NDCG@20]{Grid with 18 subplots consisting of line plots. The x axis represents the corresponding text window, the y axis contains the NDCG for a top-20 recommendation, and the six colored lines show the results obtained for each pair of embedding algorithm and state representation strategy. Results vary depending on the dataset and algorithm, with no clear consensus of superiority.}
\end{figure*}

The results of our experiments are shown in Figure~\ref{fig:ndcg_results}, which present the accumulated NDCG@20 across the 10 evaluation windows in the test set for models LinUCB, LinGreedy, and LinTS, respectively. Each line represents a pair of an embedding-based algorithm---ALS or BPR---and an aggregation function to build the context representation, as explained in Section~\ref{sec:emb-state}, which are \textit{``user''} for the user embeddings, \textit{``item mean''} for the average embedding, and \textit{``item concat''} for the concatenated embeddings. A condensed overview is also presented in Tables~\ref{tab:als-ndcg} and~\ref{tab:bpr-ndcg}, reporting the final aggregated scores when using ALS and BPR embeddings, respectively. Darker cells indicate better performance, and the best score in each column is highlighted in \textbf{bold}. For transparency and reproducibility, our source code is available at \url{https://github.com/UFSCar-LaSID/bandits_blind_spot}.

When analyzing the results, we observe that the outcomes vary drastically depending on both the embedding algorithm and the aggregation function used to construct the embedding-based states. Combinations of algorithms and aggregation strategies that achieve high scores on specific datasets can underperform on others, even within the same domain, as exemplified by the concatenated BPR embeddings in the Amazon Beauty and Amazon Books datasets.

Regarding \textbf{RQ1}, different embedding models lead to significant variations in performance, yet no single algorithm consistently outperforms the others. For instance, ALS achieved up to 22$\times$ better results than BPR on the BestBuy dataset, whereas the ranking-based BPR model outperformed ALS by a factor of 47$\times$ in specific scenarios on the MovieLens-100k dataset. Comparing the two models while holding the CMAB algorithm and aggregation function fixed shows that ALS attained the best overall score in 61\% of the experiments, with BPR thriving in the remaining 39\%. This alternating dominance suggests that although some algorithms can offer substantial gains in certain cases, their relative performance tends to vary across datasets and contexts.

We observe a similar pattern for \textbf{RQ2}. Using the same embeddings but varying the method of combining and utilizing them within the bandit recommenders leads to significant different outcomes. As an example, one might initially expect that relying solely on the static user embedding, unchanged throughout the entire online experiment, would yield poor performance. Although this strategy is generally the worst among BPR embedding approaches, as shown in Table~\ref{tab:bpr-ndcg}, it achieved the best and second-best results on the BestBuy dataset and even outperformed the item concatenation strategy for both BPR and ALS embeddings for MovieLens-100k.

\begin{table*}
\centering
\caption{ALS-based Normalized discounted cumulative gain in a top-20 recommendation task (NDCG@20).}
\label{tab:als-ndcg}
\resizebox{1.0\textwidth} {!}{ \begin{tabular}{l|ccc|ccc|ccc|}
\hline
 & \multicolumn{3}{c|}{Item mean} & \multicolumn{3}{c|}{Item concat} & \multicolumn{3}{c|}{User}\\ & LinUCB & LinGreedy & LinTS & LinUCB & LinGreedy & LinTS & LinUCB & LinGreedy & LinTS \\
\hline
Amazon Beauty & \cellcolor[gray]{0.5575886302462387}0.00577 & \cellcolor[gray]{0.5015823963379868}0.00589 & \cellcolor[gray]{0.5}\textbf{0.00590} & \cellcolor[gray]{0.853526506889438}0.00514 & \cellcolor[gray]{0.8988256115933084}0.00504 & \cellcolor[gray]{0.8299266452263558}0.00519 & \cellcolor[gray]{0.828343312734106}0.00519 & \cellcolor[gray]{0.9500000000000002}0.00493 & \cellcolor[gray]{0.8277547284899374}0.00519 \\
Amazon Books & \cellcolor[gray]{0.9500000000000001}0.00473 & \cellcolor[gray]{0.938640180772054}0.00518 & \cellcolor[gray]{0.9385516581649674}0.00518 & \cellcolor[gray]{0.5072950241007358}0.02243 & \cellcolor[gray]{0.5052405482316029}0.02251 & \cellcolor[gray]{0.5}\textbf{0.02272} & \cellcolor[gray]{0.8860737818586237}0.00728 & \cellcolor[gray]{0.8862709347109647}0.00727 & \cellcolor[gray]{0.8847513053215523}0.00733 \\
BestBuy & \cellcolor[gray]{0.5945104255967409}0.02191 & \cellcolor[gray]{0.5957307407489004}0.02186 & \cellcolor[gray]{0.6460779112230368}0.01969 & \cellcolor[gray]{0.6170537696737757}0.02094 & \cellcolor[gray]{0.5066624225581073}0.02569 & \cellcolor[gray]{0.5}\textbf{0.02597} & \cellcolor[gray]{0.9489611564974116}0.00667 & \cellcolor[gray]{0.95}0.00663 & \cellcolor[gray]{0.9486195331641967}0.00669 \\
Delicious & \cellcolor[gray]{0.7827161372187741}0.03115 & \cellcolor[gray]{0.7772656201552586}0.03152 & \cellcolor[gray]{0.7777657674239651}0.03148 & \cellcolor[gray]{0.5336304169576224}0.04796 & \cellcolor[gray]{0.505147860870149}0.04989 & \cellcolor[gray]{0.5000000000000001}\textbf{0.05024} & \cellcolor[gray]{0.9500000000000002}0.01985 & \cellcolor[gray]{0.949955675132588}0.01986 & \cellcolor[gray]{0.9401999533885042}0.02052 \\
MovieLens-100k & \cellcolor[gray]{0.6417897214189241}0.04087 & \cellcolor[gray]{0.6470439895428586}0.04018 & \cellcolor[gray]{0.6410039926444364}0.04097 & \cellcolor[gray]{0.5}\textbf{0.05946} & \cellcolor[gray]{0.5044009701227042}0.05889 & \cellcolor[gray]{0.5000105426929613}0.05946 & \cellcolor[gray]{0.95}0.00045 & \cellcolor[gray]{0.9334503879349834}0.00262 & \cellcolor[gray]{0.9228182987321119}0.00402 \\
MovieLens-25M & \cellcolor[gray]{0.5265554896613966}0.01308 & \cellcolor[gray]{0.5140204386859011}0.01321 & \cellcolor[gray]{0.5}\textbf{0.01335} & \cellcolor[gray]{0.9500000000000001}0.00879 & \cellcolor[gray]{0.9463181340561382}0.00883 & \cellcolor[gray]{0.8241066683496413}0.01007 & \cellcolor[gray]{0.6287744620270665}0.01205 & \cellcolor[gray]{0.6062595177843555}0.01227 & \cellcolor[gray]{0.5936631319899282}0.01240 \\
\hline
\end{tabular} }
\end{table*}

Contrary to prior assumptions that concatenation would be the superior approach (due to preserving the ordering of item consumption), the other strategies surpassed concatenation in 41\% of the experiments. These findings highlight that, beyond exploring different embedding algorithms for potential improvements (\textbf{RQ1}), it is equally critical to evaluate the optimal method for leveraging these embeddings in constructing the state representation (\textbf{RQ2}).

\begin{table*}
\centering
\caption{BPR-based Normalized discounted cumulative gain in a top-20 recommendation task (NDCG@20).}
\label{tab:bpr-ndcg}
\resizebox{1.0\textwidth} {!}{ \begin{tabular}{l|ccc|ccc|ccc|}
\hline
 & \multicolumn{3}{c|}{Item mean} & \multicolumn{3}{c|}{Item concat} & \multicolumn{3}{c|}{User}\\ & LinUCB & LinGreedy & LinTS & LinUCB & LinGreedy & LinTS & LinUCB & LinGreedy & LinTS \\
\hline
Amazon Beauty & \cellcolor[gray]{0.7398028752572645}0.00777 & \cellcolor[gray]{0.7342972085555113}0.00779 & \cellcolor[gray]{0.6917525006028775}0.00795 & \cellcolor[gray]{0.5072667824675698}0.00867 & \cellcolor[gray]{0.5430616645704016}0.00853 & \cellcolor[gray]{0.5}\textbf{0.00870} & \cellcolor[gray]{0.9499999999999997}0.00695 & \cellcolor[gray]{0.9362475466932072}0.00701 & \cellcolor[gray]{0.8303910202284677}0.00742 \\
Amazon Books & \cellcolor[gray]{0.8215381383713567}0.00123 & \cellcolor[gray]{0.8165756930320673}0.00127 & \cellcolor[gray]{0.8130913585992438}0.00129 & \cellcolor[gray]{0.519126047578711}0.00332 & \cellcolor[gray]{0.5044593035324104}0.00342 & \cellcolor[gray]{0.5}\textbf{0.00345} & \cellcolor[gray]{0.95}0.00035 & \cellcolor[gray]{0.940776706381126}0.00041 & \cellcolor[gray]{0.8256381097243101}0.00121 \\
BestBuy & \cellcolor[gray]{0.95}0.00098 & \cellcolor[gray]{0.940611927968562}0.00110 & \cellcolor[gray]{0.7446722839140787}0.00379 & \cellcolor[gray]{0.8934184124021385}0.00175 & \cellcolor[gray]{0.8703612026837975}0.00207 & \cellcolor[gray]{0.7444537990356559}0.00380 & \cellcolor[gray]{0.9253911504431689}0.00131 & \cellcolor[gray]{0.6694935182826631}0.00483 & \cellcolor[gray]{0.5}\textbf{0.00715} \\
Delicious & \cellcolor[gray]{0.7546436415650826}0.07363 & \cellcolor[gray]{0.7252151377029477}0.08276 & \cellcolor[gray]{0.734000884263416}0.08003 & \cellcolor[gray]{0.5}\textbf{0.15259} & \cellcolor[gray]{0.5426247251932896}0.13938 & \cellcolor[gray]{0.5402151363526768}0.14012 & \cellcolor[gray]{0.95}0.01306 & \cellcolor[gray]{0.8981878524854006}0.02912 & \cellcolor[gray]{0.9023216385998892}0.02784 \\
MovieLens-100k & \cellcolor[gray]{0.5067783674500601}0.03079 & \cellcolor[gray]{0.7327760570427243}0.02388 & \cellcolor[gray]{0.5}\textbf{0.03100} & \cellcolor[gray]{0.5322988591718236}0.03001 & \cellcolor[gray]{0.9499999999999998}0.01723 & \cellcolor[gray]{0.7125023067588517}0.0245 & \cellcolor[gray]{0.81842806453795}0.02126 & \cellcolor[gray]{0.8118334700279838}0.02146 & \cellcolor[gray]{0.7958612474436153}0.02195 \\
MovieLens-25M & \cellcolor[gray]{0.5137415391992798}0.01005 & \cellcolor[gray]{0.502709576279386}0.01027 & \cellcolor[gray]{0.5}\textbf{0.01032} & \cellcolor[gray]{0.95}0.00147 & \cellcolor[gray]{0.9442899426208536}0.00158 & \cellcolor[gray]{0.7885510039273446}0.00464 & \cellcolor[gray]{0.7338193121080714}0.00572 & \cellcolor[gray]{0.6990601383456774}0.0064 & \cellcolor[gray]{0.5578717212299631}0.00918 \\
\hline
\end{tabular} }
\end{table*}

To further address the first two research questions, we conducted a Friedman test on the results presented in Tables~\ref{tab:als-ndcg} and~\ref{tab:bpr-ndcg} to assess statistical differences among models. First, we compared the results obtained with ALS and BPR embeddings (\textbf{RQ1}). The test revealed no statistically significant difference ($\chi^2_r = 2.67$, $p = 0.102$). Next, we applied the same procedure to compare the three aggregation strategies (\textbf{RQ2}). At a 99\% confidence level, the test indicated significant differences across methods ($\chi^2_r = 16.26$, $p = 0.00029$). A post-hoc Nemenyi test was then performed to identify pairwise differences. With a critical difference of $0.5522$, results show that using the \textit{Item-averaged} and \textit{Concatenated item} embeddings are statistically superior to the \textit{User} embeddings, with a 95\% confidence, while no significant difference was observed between the two former methods.

Turning to \textbf{RQ3}, our results demonstrate that the quality of the state representation often exerts a greater influence on recommendation performance than the choice of the bandit algorithm itself. As summarized in Tables~\ref{tab:als-ndcg} and~\ref{tab:bpr-ndcg}, the performance of different CMAB algorithms using the same state representation is generally very similar, with only a few notable improvements in specific cases, such as the Delicious and MovieLens-100k datasets when employing the item concatenation strategy. Typically, differences among bandit algorithms appear only in the fourth decimal place. Across all datasets, switching from a lower-quality embedding representation to a more effective aggregation method or embedding model resulted in substantially larger performance gains than changing the underlying bandit algorithm.
 
In summary, our experiments highlight that both the choice of embedding algorithm and the strategy used to aggregate embeddings significantly impact recommendation performance. No single embedding model or aggregation method consistently dominates across datasets, underscoring the need for tailored evaluation. Moreover, the quality of the state representation generally has a stronger effect on the results than the specific bandit algorithm employed, suggesting that future work should prioritize optimizing embedding-based context representations alongside algorithmic improvements. These findings reveal critical considerations for the design of effective CMAB recommenders.

\section{Conclusions and Future Work}

In this work, we investigated the impact of state representation on the performance of contextual multi-armed bandit (CMAB) recommendation models. Through a comprehensive set of experiments, we evaluated multiple embedding models, different strategies for leveraging embeddings (e.g., user-based, aggregated representations), and contrasted their influence with that of bandit algorithm selection. Our results provide evidence that the choice of state representation can play a role as critical as, and in most cases even more decisive than, the choice of the CMAB algorithm itself. By highlighting this relationship, we aim to bring attention to an often overlooked aspect of MAB-based recommendation: the design and evaluation of the state representation.

Our experiments demonstrate that variations in the state representation can yield improvements up to 47$\times$ larger compared to changes in the choice of the CMAB algorithm, which typically influenced results only at the fourth decimal place. Moreover, we found no definitive consensus on the optimal state representation strategy; the two tested embedding-based algorithms alternated in achieving the best scores across different datasets and settings. These findings highlight the critical importance of carefully designing and selecting embedding-based context representations, suggesting that future advances in contextual bandit recommenders should prioritize state representation quality as much as, if not more than, the development of new bandit algorithms.

As future work, we plan to extend our study in several directions. First, we intend to explore different SOTA neural embedding-based algorithms, such as RecVAE~\cite{ShenbinRecVAE2019} and Neural Collaborative Filtering~\cite{HeNeural2017}. Second, we will evaluate in our experimental protocol SOTA deep learning incremental recommender systems, such as BSARec~\cite{ShinBsarec2024}, GRU4RecCPR~\cite{ChangGRU4RecCpr2024}, and FEARec~\cite{DuFEARec2023}. We also plan to analyze the influence of embedding hyperparameter selection on CMAB outcomes, aiming to quantify its role in shaping recommendation quality. Lastly, we will investigate the differences between static embeddings and incrementally learned embeddings to determine how dynamic adaptation impacts long-term performance.

\begin{acks}
This study was financed, in part, by the Brazilian Agencies CNPq (grant \#311867/2023-5), CAPES (Finance Code 88887.854357/2023-00), and FAPESP (Process Numbers \#2021/14591-7, \#2023/00158-5, and \#2024/15919-4).
\end{acks}

\balance
\bibliographystyle{ACM-Reference-Format}
\bibliography{references}


\end{document}